\documentclass[%
 aip,
 amsmath,amssymb,
 reprint,%
]{revtex4-1}

\usepackage{graphicx}
\usepackage{dcolumn}
\usepackage{bm}

\usepackage[utf8]{inputenc}
\usepackage[T1]{fontenc}
\usepackage{mathptmx}
\usepackage{etoolbox}
\usepackage{textcomp}
\usepackage[separate-uncertainty=true]{siunitx}

\makeatletter
\def\@email#1#2{%
 \endgroup
 \patchcmd{\titleblock@produce}
  {\frontmatter@RRAPformat}
  {\frontmatter@RRAPformat{\produce@RRAP{*#1\href{mailto:#2}{#2}}}\frontmatter@RRAPformat}
  {}{}
}%
\makeatother
\begin{document}

\newcommand{\mycite}[1]{[\onlinecite{#1}]}

\preprint{AIP/123-QED}

\title[]{Phonon-dominated thermal conductivity of a superconducting amorphous MoSi thin film}

\author{Jacopo Piantanida Chiesa}

\author{Tom Q. Post}

\author{Martin P. van Exter}

\author{Michiel J.A. de Dood*}
\affiliation{Huygens-Kamerlingh Onnes Laboratory, Leiden University, Niels Bohrweg 2, Leiden 2333 CA, The Netherlands}
\email{dood@physics.leidenuniv.nl}

\date{19 October 2026}

\begin{abstract}
We use the differential $3\omega$ method to directly measure the temperature dependent thermal conductivity for $\sim90$ nm Mo$_{0.78}$Si$_{0.22}$ thin films. The measurements are performed in a closed-cycle cryostat to access the temperature range between $50$ K and $300$ K. The MoSi thin films are capped with $\text{Al}_2\text{O}_3$ and probed with a Pt wire. The results show that the thermal conductivity does not satisfy the Wiedemann-Franz law. This implies that thermal transport in MoSi is not dominated by electrons but by phonon-like excitations, contrary to what is assumed in many electro-thermal models of superconducting devices. We argue that this property might be related to the weak localization visible in the resistivity as a function of temperature curves of the superconductor. 
\end{abstract}

\maketitle

Thin film superconductors play a central role in many quantum technologies. Rapid Single Flux Quantum circuits are at the heart of an entire quantum computing archetype\cite{krylov2022}, Superconducting Nanowire Single Photon Detectors (SNSPDs) can reach near-100\% detection efficiencies in the telecom and visible range\cite{chang2021}, Transition Edge Sensors are slower but provide reliable photon number resolution\cite{fukuda2008}, while Microwave Kinetic Inductance Detectors are promising for large single-photon cameras for space applications\cite{visser2021}. 

The working principle of many of these devices relies on out-of-equilibrium superconductivity, including a temporary local breaking of the superconductive state. Any current carrying structure with a sudden increase in resistance will be subjected to Joule heating. An independent study of the thermal properties of the relevant superconducting materials in their normal state thus provides additional important insights into their dynamical behavior and thermal reset\cite{yin2024}. 

We consider the example of SNSPDs, where the absorption of a single photon in a current-biased superconducting nanowire provides enough energy to locally break superconductivity\cite{goltsman2001}. An SNSPD detects a photon as a voltage pulse across a shunted wire to prevent latching\cite{yang2007}, and the detector reset time is ultimately limited by thermal physics. Multiple electro-thermal models of this system have been proposed, but most of them assume a Wiedemann-Franz law to describe the superconductor's thermal conductivity in its normal state\cite{skocpol1974, yang2007}. This implies that thermal transport is entirely dominated by electrons and that the thermal conductivity is inversely proportional to the electrical resistivity: $\kappa(T)\propto T/\rho(T)$. A similar assumption is made in the thermal analysis of rapid single flux quantum circuits\cite{krylov2022, mitrovic2022}. 

SNSPDs make use of disordered superconducting materials with low carier density (\textit{e.g.} NbTiN\cite{zichi2019}, NbN\cite{lita2015}, MoSi\cite{lita2015}, WSi\cite{lita2015}...) whose normal-state resistivity increases with decreasing temperature due to weak localization effects\cite{gantmakher2011}. Given the unusual nature of the electronic behavior in these materials, we believe that the assumption that their thermal transport is fully electron-dominated is a rather strong one. A direct measurement of the temperature dependent thermal conductivity for these superconductors is therefore in order. 

In this work we use the differential $3\omega$ method to measure the thermal conductivity of amorphous superconducting MoSi thin films. In section \ref{sec:threeOmegaMethod} we introduce the $3\omega$ technique for thermal conductivity measurements, in section \ref{sec:sample_prep} we discuss sample preparation and characterization of the thin films. Section \ref{sec:ThreeOmegaMeasurements} describes the details of the $3\omega$ measurements for our samples and section \ref{sec:discussion} discusses the details of the measurement results, arguing the limitations of the Wiedemann-Franz law in this context. Finally, section \ref{sec:conclusion} presents the implication of our results and the influence they have in thermal modeling of superconducting devices.  

\section{\label{sec:threeOmegaMethod}The 3$\omega$ method}

The $3\omega$ method, first proposed by Cahill\cite{cahill1987, cahill1994}, measures the thermal conductivity of a material by using a metal wire both as a heater and as a thermometer. The wire is current driven at frequency $\omega$ to create an output voltage at frequency $3 \omega$ from which the thermal conductivity can be extracted. For a sample of thickness $d_s$ and a wire of width $2a$ and length $l$, one can find an expression for the temperature oscillations $\Delta T$ in the sample\cite{cahill1987}: 
\begin{equation}
    \Delta T(t) \approx \frac{P(t)}{2 \pi \kappa_s l} \left[\ln{\frac{D}{2a^2}}+2\eta-j\frac{\pi}{2}-\ln{\omega}\right] ~, 
    \label{eq:deltaT}
\end{equation}
valid in the limit $a \ll \sqrt{D/2\omega} \ll \{d_s,l\}$. Here $D$ is the heat diffusion coefficient of the specimen, $P(t)=P_0e^{2j\omega t}$ is the Joule power generated by the current, $\kappa_s$ is the thermal conductivity of the specimen, $j$ is the imaginary unit and $\eta\approx1$ is a numerical constant. In practice, it is often more convenient to work with the thermal resistance $Z = \Delta T /P$, that normalizes this temperature difference by making it power independent\cite{dames2005}.

The voltage across the wire follows from Ohm's law: 
\begin{equation}
    V(t) = I_0R_0 e^{j\omega t} + \frac{I_0 P_0}{2 \pi \kappa_s l } \frac{dR}{dT}\left[ \ln{\frac{D}{2a^2}}+2\eta-j\frac{\pi}{2} -\ln{\omega}\right] e ^{3j\omega t} ~,
    \label{eq:3omegaVolt}
\end{equation}
which contains a higher harmonic at frequency $3\omega$. Here $R(T) \approx R_0 + \frac{dR}{dT} \Delta T$ is the wire's resistance, while $I_0$ is the amplitude of the bias current. The thermal resistance of the sample follows from the amplitude of this $3 \omega$ component $Z=V_{3\omega}/(P_0I_0\frac{dR}{dT})$. 

The addition of a thin film of thickness $d_f\ll a$ gives an additional frequency-independent thermal resistance $Z_f$ \cite{cahill1994} 
\begin{equation}
    Z_f = \frac{1}{\kappa_f l} \left(\frac{d_f}{2a}\right) ~,
    \label{eq:Zf}
\end{equation}
where $\kappa_f$ is the thermal conductivity of the thin film. We employ the differential $3\omega$ method\cite{dames2012} and create two separate sample stacks deposited on the same chip: one includes the thin film of interest and the other is a reference without it. The in-phase component of the $3\omega$ signal for the stack with the thin film will have a positive offset when compared to the reference stack. A measurement of this offset gives access to $Z_f$ and therefore to the thermal conductivity of the thin film. The differential method can also be applied to capped thin films, as long as the capping material is included in the reference stack. This allows to measure the thermal conductivity of electrically conductive thin films without worrying about current leakage from the measurement wire\cite{song1999}.  

\begin{figure}
\includegraphics{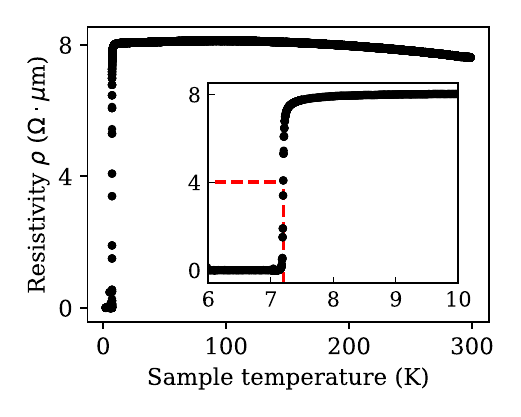}
\caption{\label{fig:RTcurve} Resistivity as a function of temperature for a \qty{91 \pm 1}{\nm} thick Mo$_{0.78}$Si$_{0.22}$ film capped with $\simeq \qty{100}{\nm}$ of $\text{Al}_2\text{O}_3$. The critical temperature $T_C = \qty{7.2 \pm 0.1}{\kelvin}$ is calculated as the temperature where the resistivity reaches half of the value at \qty{10}{\kelvin} (inset, dashed lines). }
\end{figure}

\section{\label{sec:sample_prep}Sample preparation and characterization}
\begin{figure*}
\includegraphics{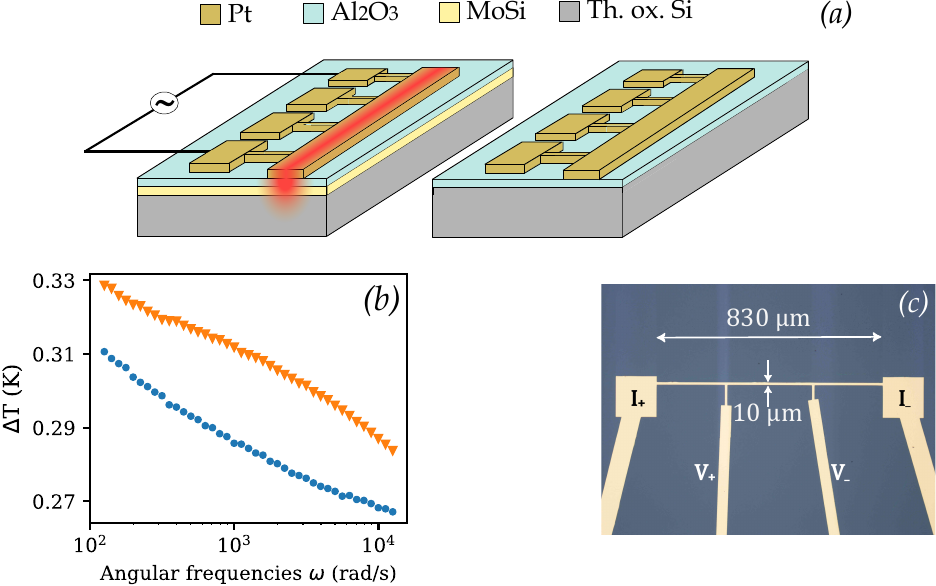}
\caption{\label{fig:fig2}The $3\omega$ method applied to thin film MoSi. (a) Simplified schematic of the $3\omega$ sample including the superconducting MoSi layer (left) and reference (right). Each wire is probed with a four-point-probe measurement and the two stacks are grown on the same chip. (b) Measured $3\omega$ signal $\Delta T$ as a function of $\omega$ on a logarithmic scale  at $T\simeq\qty{260}{\kelvin}$. The data for the stack including the MoSi thin film (orange triangles) are clearly above the data  for the reference stack without the superconductor (blue circles). The offset between the two lines corresponds to a thermal conductivity of MoSi $\kappa \simeq \qty[per-mode=fraction]{4.8}{\watt \per \meter \per \kelvin}$. (c) Optical microscope image of a Pt wire used for four-point-probe measurement in a $3\omega$ sample. }. 
\end{figure*}

Figure \ref{fig:fig2}(a) shows schematics of the samples used in this work. The material stacks were deposited via RF diode sputtering on a \qty{525}{\um} thick Si(100) wafer ($\rho_\text{Si} =0.001-0.005~ \Omega \cdot \text{cm}$) with \qty{300}{\nm} of wet thermal oxide on top. All materials were deposited in an Ar plasma at \qty{7.5e-3}{\milli \bar} (base pressure $<\qty{e-5}{\milli \bar}$) with sputtering targets biased at \qty{1}{\kilo \volt} . The amorphous $\text{Mo}_x\text{Si}_{1-x}$ layer was deposited from a composite target and results in a film with $x\simeq 0.78$, as obtained from Energy Dispersive X-ray Spectroscopy.

The stack and the reference (see  \ref{fig:fig2}(a)) are grown on the same $10\times10$ mm substrate. The substrate is partially masked and $\sim \qty{90}{\nm}$ of MoSi is deposited on a $\sim$$2.5\times10$ mm stripe. The mask is then removed, taking care to leave the sample out of vacuum for less than five minutes, and a capping layer of $\sim100$ nm of $\text{Al}_2\text{O}_3$ is deposited to electrically insulate the sample. $\text{Al}_2\text{O}_3$ was chosen because of its relatively high thermal conductivity. The $\sim \qty{180}{\nm}$ thick Pt wires for four-point-probe measurements were defined on top using a combination of electron beam lithography and subsequent liftoff techniques, resulting in the wire geometry shown in figure \ref{fig:fig2}(c).

The thickness of the layers was determined by Atomic Force Microscopy as $d_f = \qty{91 \pm 1}{\nm}$ and $d_w = \qty{182 \pm 1}{\nm}$ for the MoSi layer and Pt wire, respectively. The thickness of the $\text{Al}_2\text{O}_3$ layer was estimated as $\qty{100 \pm 8}{\nm}$ from the known sputter rate of our target.  

To determine the resistivity and superconducting transition of the MoSi film, a sample with the same thickness and composition was used and patterned with a four-point-probe  square van-der-Pauw geometry following the procedure described by NIST\cite{nistVdP}. This sample was capped with the same $\text{Al}_2\text{O}_3$ layer, to account for potential influence of capping on the superconducting critical temperature of the film. 
The temperature dependent resistivity was measured in a pulse tube cryostat (Oxford Instruments Teslatron), reaching a base temperature of \qty{1.4}{\kelvin}. The resistances were calculated using a \qty{10}{\micro \ampere} probe current provided by a Keithley 6221 current source, while voltages were measured with a Keithley 2182A voltmeter. 

Figure \ref{fig:RTcurve} shows a resistivity vs temperature curve for our superconductive Mo$_{0.78}$Si$_{0.22}$ thin film. The critical transition happens at $T_C=\qty{7.2 \pm 0.1 }{\kelvin}$, defined as the temperature where the resistivity is half of its value at \qty{10}{\kelvin} (see inset, dashed lines). The Residual Resistance Ratio RRR$=\rho_{300\text{K}}/\rho_{10\text{K}}=0.95$ agrees with the expectations of weak localization in the disordered metallic state\cite{gantmakher2011}.  The Wiedemann-Franz law allows to predict the electronic contribution to resistivity\cite{schwiete2014}: 
\begin{equation}
    \kappa_e (T) = \frac{LT}{\rho(T)} \simeq \frac{LT}{\rho_{10\text{K}}} ~, 
    \label{eq:WF}
\end{equation}
where $L=\qty{2.44e-8}{\volt^2\per \kelvin^2}$ is the Lorenz number. The last approximation is justified due to the negligible variations of the resistivity in the normal state.

\section{\label{sec:ThreeOmegaMeasurements}Results: $3\omega$ measurements}
The $3\omega$ measurements were performed using a Stanford Research System SR830 Digital Lock-in Amplifier as the AC source and detector for the experiment. The four-point-probe configuration (fig.\ref{fig:fig2}(c)) factors out the effects of measurement cables and contact resistance. For each of the wires, the raw voltage signal is first fed into a printed-circuit board (PCB) that performs active subtraction of the $1\omega$ component (see supplemental info). This subtraction limits overload issues on the lock-in amplifier and removes any spurious higher order harmonics present in the driving signal.The active components on the PCB enable the subtraction at different temperatures and thus correct for changes in the resistance of the Pt wire. The output of the PCB is connected to the lock-in amplifier to detect the $3\omega$ component.

Before cooldown every wire was probed at room temperature at a fixed frequency and varying AC amplitude, to verify that $V_{3\omega} \propto V_{1\omega}^3$ in agreement with equation (\ref{eq:3omegaVolt}). Only samples that passed this check were inserted into the cryostat. The cryostat controls the temperature in the range \qtyrange{1.5}{300}{\kelvin} to an accuracy of \qty{0.1}{\kelvin} with the sample immersed in He exchange gas. The  $3\omega$ measurements were performed in the temperature range \qtyrange{50}{295}{\kelvin}. Below \qty{50}{\kelvin} the temperature coefficient of resistance for Pt ($\frac{1}{R}\frac{\mathrm{d}R}{\mathrm{d}T}$) becomes too small for the $3\omega$ method.\\

Figure \ref{fig:fig2}(b) shows a representative measurement of the $3\omega$ signal converted to $\Delta T$ as a function of angular frequency at $T\simeq \qty{260}{\kelvin}$. The measurements for the MoSi stack (orange triangles) are clearly above the data for the reference stack (blue circles). The temperature increase $\Delta T$ is derived from the $3\omega$ signal and can be understood as a peak-to-peak temperature difference between the top and bottom of the sample during the measurement. While the $\Delta T$ representation gives immediate physical insight, the use of the thermal resistance $Z$ defined via Eq.(\ref{eq:3omegaVolt}) corrects for small differences between the two wires (see supplemental info). The data in \ref{fig:fig2}(b) are converted to $Z(\ln{\omega})$ and straight lines are fit to this data to obtain $Z_f$. These data are then converted to a thermal conductivity using Eq. (\ref{eq:Zf}). 

\section{\label{sec:discussion}Discussion}
\begin{figure}
\includegraphics{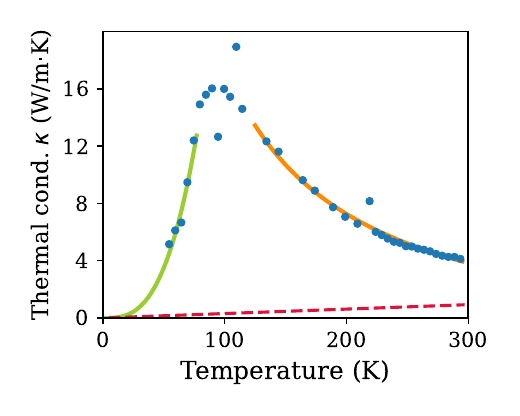}
\caption{\label{fig:kappas} Thermal conductivity as a function of temperature for a Mo$_{0.78}$Si$_{0.22}$ film of thickness $d=\qty{91}{\nm}$, measured between \qty{50}{\kelvin} and \qty{300}{\kelvin}. For high temperature data the fit function is $\kappa_\text{high}(T) = \frac{\left(\qty[per-mode=fraction]{2060}{\watt \per \meter}\right)}{T} -\left(\qty[per-mode=fraction]{3.0}{\watt \per \meter \per \kelvin}\right)$ (orange line), while for low temperatures we fit with $\kappa_\text{low} (T) = \left(2.8\cdot10^{-5} \frac{\mathrm{W}}{\mathrm{\mathrm{m } \mathrm{K}^4 }}\right) ~ T^3$ (green line). Both behaviors agree with the expectations for a thermal conductivity dominated by phonon-like excitations. The expected contribution from the Wiedemann-Franz law to $\kappa$ based on the measured resistivity is insignificant and is shown by the red dashed line.}
\end{figure}

Figure \ref{fig:kappas} shows the results for the temperature dependent thermal conductivity measurements of thin film MoSi, from the samples described in section \ref{sec:sample_prep}. 
The thermal conductivity first increases with decreasing temperature, reaches a maximum around \qty{95}{\kelvin} and then decreases with decreasing temperature.
The dashed red line shows the prediction from the Wiedemann-Franz law (\ref{eq:WF}) based on the resistivity curve in Fig.~\ref{fig:RTcurve}. The measured temperature dependence of the thermal conductivity clearly follows a different trend. Moreover, the electronic contribution to the thermal conductivity is small compared to other contributions. The high temperature data show a $1/T$ dependence fitted by the orange line  in Fig.~\ref{fig:kappas}. The low temperature data follow a $T^3$ law as indicated by the green line.

The $T^3$ dependence at low temperature can be explained in terms of kinetic theory and occupation of the phononic bands. The $1/T$ dependence at high temperatures, also phononic in nature, comes from the decrease in the scattering length with increasing temperature\cite{ziman67}. It is not clear if this theory can be applied to amorphous materials such as the MoSi studied here because phonons are less well-defined. However, the same temperature dependence can be obtained also in continuous models that do not assume any lattice structure\cite{debye1914}. In either case, our result is indicative of a thermal conductivity that is dominated by phonon-like excitations, exemplifying the need to go beyond the Wiedemann-Franz law for thermal simulations of MoSi. 

According to Makinson\cite{makinson1938}, an ideal system to measure the lattice thermal conductivity of a metal would be an alloy with high resistivity to reduce the scattering length for electrons but not for phonons. 
It is possible to show that weak localization effects in MoSi can be explained by strong electron-electron scattering with a corresponding short scattering length~\cite{gantmakher2011}. We are not aware of additional effects that would significantly reduce the phononic scattering length, and hence interpret our results in terms of phonon-dominated thermal conductivity. 

Given that the great majority of SNSPD-relevant materials show signs of weak localization effects in their temperature dependent resistivity, we believe that direct measurements of the thermal conductivity for these materials are necessary. 

\section{\label{sec:conclusion}Conclusion}

We presented a direct measurement of the thermal conductivity of thin fim MoSi using the differential $3\omega$ method. Measurements were performed on a \qty{91 \pm 1}{\nm} thick Mo$_{0.78}$Si$_{0.22}$ film capped with $\text{Al}_2\text{O}_3$, probed with a Pt wire in the temperature range between \qty{50}{\kelvin} and \qty{300}{\kelvin}. Parasitic effects are avoided by employing a four-point-probe measurement configuration, using an active $1\omega$ subtraction PCB, and taking advantage of the differential nature of the method.

The temperature dependence of the thermal conductivity shows a lack of electronic contributions and indicates that thermal transport in MoSi is dominated by phonon-like excitations. Electrothermal models that rely solely on the Wiedemann-Franz law to simulate thermal transport in MoSi assume that heat transport is dominated by electron transport and are thus incomplete. We suggest that the phonon dominated behavior is related to the short scattering length for electrons in the amorphous MoSi material, which shows clear signs of weak-localization effects in the resistivity. Other weakly localized systems that are commonly used in SNSPDs due to their low carrier density could show similar behavior. An extension of  these measurements to comparable materials could therefore give additional insight on heat transport in these material systems. 

\begin{acknowledgments}
The authors acknowledge fruitful scientific discussions with Christopher Dames, UC Berkeley on the use of the 3$\omega$ method. We thank Shuhul Mujoo, Caltech for the contribution at initial stage of this work. This research was made possible by financial support from the Dutch Research Council (NWO).
\end{acknowledgments}

\section*{Data Availability Statement}

The data that support the findings of this study are available from the corresponding author upon reasonable request.

\nocite{*}
\bibliography{bibliography}

\end{document}